\documentclass[a4paper,11pt]{article}
\pdfoutput=1
\usepackage{jcappub}
\usepackage{amssymb}
\usepackage{graphicx}
\usepackage{amsmath}
\usepackage{hyperref}

\title{\boldmath Constraints on non-minimal coupling from quantum cosmology}

\author[a]{Shao-Jiang Wang,}
\author[a]{Masaki Yamada,}
\author[a]{Alexander Vilenkin}

\affiliation[a]{Tufts Institute of Cosmology, Department of Physics and Astronomy, Tufts University, 574 Boston Avenue, Medford, Massachusetts 02155, USA}

\emailAdd{schwang@cosmos.phy.tufts.edu}
\emailAdd{Masaki.Yamada@tufts.edu}
\emailAdd{vilenkin@cosmos.phy.tufts.edu}

\abstract{Quantum cosmology is investigated in a de Sitter minisuperspace model with a quantized scalar field non-minimally coupled to curvature.  Quantum states of the scalar field must satisfy the regularity condition, which requires that the probability of field fluctuations should not increase with their amplitude.  We show that this condition imposes constraints on the allowed values of the curvature coupling parameter $\xi$.  This is a surprising result, since the field dynamics depends only on the combination $m^2+\xi R$, where $m$ is the field mass and $R = \mathrm{const}$ is the curvature, and does not depend on $\xi$ separately.}

\begin{document}
\maketitle
\flushbottom

\section{Introduction}\label{sec:int}

In quantum cosmology the wave function of the universe $\Psi(g,\phi)$ is defined on the space of all 3-geometries ($g$) and matter field configurations $(\phi)$, called superspace.  This wave function can be found by solving the Wheeler-DeWitt (WDW) equation \cite{DeWitt:1967yk}
\begin{align}
\mathcal{H}\Psi=0,
\label{WhDW}
\end{align}
where $\mathcal{H}$ is the Hamiltonian operator. (For a review of quantum cosmology and references to the early literature see, e.g., \cite{Halliwell:1990uy}.) The solution of Eq.~(\ref{WhDW}) depends on one's choice of boundary conditions for $\Psi$, which have to be postulated as an independent physical law. The two widely studied proposals for this law of boundary conditions are the no-boundary proposal of Hartle and Hawking \cite{Hartle:1983ai,Hawking:1983hj} and the tunneling proposal \cite{Vilenkin:1986cy,Vilenkin:1987kf}.\footnote{For early discussion of the tunneling and related proposals see also \cite{Linde:1983mx,Rubakov:1984bh,Vilenkin:1984wp,Zeldovich:1984vk}.} The wave function of the universe can also be defined as a path integral. In recent years there has been a heated debate as to which definitions of the path integral are mathematically consistent and which (real or complex) paths should be included in the integral \cite{Feldbrugge:2017kzv,Feldbrugge:2017fcc,Feldbrugge:2017mbc,Feldbrugge:2018gin,Lehners:2018eeo,DiazDorronsoro:2017hti,DiazDorronsoro:2018wro,Vilenkin:2018dch,Vilenkin:2018oja}. In the present paper we shall use the WDW formalism of quantum cosmology and will not be concerned with this debate.

Much of the work in quantum cosmology has focused on de Sitter minisuperspace models, where the geometry is restricted to that of a closed, homogeneous and isotropic universe and the matter content is restricted to a cosmological constant and some quantum fields which are treated as small perturbations. The wave function can then be represented as a superposition of terms of the form
\begin{align}
\mathrm{e}^{iS(a)}\mathrm{e}^{-\sum\limits_n R_n(a)\phi_n^2} ,
\end{align}
where $a$ is the scale factor, which is assumed to be a semiclassical variable, $S(a)$ is the classical action, which can be imaginary in the classically forbidden region, and $\phi_n$ are the amplitudes of scalar field modes.  A physically acceptable quantum state should satisfy the condition
\begin{align}
\mathrm{Re}R_n(a)\geq0,
\end{align}
so that the probability of quantum fluctuations does not grow with their amplitude. We shall refer to this as the regularity condition. It will be our main focus in this paper.

Minisuperspace models of the kind outlined above have been extensively studied for the cases of scalar fields minimally and conformally coupled to scalar curvature, while there are only a few works on the cases of other curvature coupling (for example, \cite{Fakir:1990zi}). However, there is no convincing reason, apart from simplicity, that nature must choose such special couplings. In fact, it is well known that a non-minimal coupling for a scalar field is generally induced at one-loop order in the interacting theory, even if it is absent at the tree level \cite{Freedman:1974gs}. Non-minimal coupling plays a crucial role the in Higgs inflation model \cite{Spokoiny:1984bd,Bezrukov:2007ep} and Higgs stability problem (see \cite{Espinosa:2018mfn,Markkanen:2018pdo} for a brief review). On the other hand, current observational constraints on non-minimal coupling are quite loose, for example $|\xi|\lesssim10^{15}$ from the LHC's result \cite{Atkins:2012yn}. 

In this paper, we shall explore de Sitter minisuperspace quantum cosmology with a non-minimally coupled scalar field.\footnote{See also \cite{Barvinsky:2009jd,Barvinsky:2015uxa} for an alternative treatment of the non-minimal coupling in the context of quantum cosmology using a microcanonical density matrix state of the Universe.} We find that acceptable quantum states for the scalar field exist only when the curvature coupling parameter $\xi$ is restricted to a certain range. The range is different for the tunneling and no-boundary wave functions. This result is surprising, since the dynamics of the scalar field depends only on the effective mass, $m_\mathrm{eff}^2=m^2+\xi R$, where $m$ is the field mass and $R$ is the curvature, and does not depend on $\xi$ separately.

The outline of this paper is as follows: In section \ref{sec:WDW} the minisuperspace model with a non-minimally coupled scalar field is introduced. In section \ref{sec:constraint}, constraints on the curvature coupling parameter are derived from the regularity condition for both the tunneling and no-boundary wave functions. In section \ref{sec:con} our results are briefly summarized and discussed.

\section{The model}\label{sec:WDW}

We consider a massive scalar field $\phi$ non-minimally coupled to the Ricci scalar, with the spacetime metric of a closed Friedmann-Lemaitre-Robertson-Walker form 
\begin{align}
\mathrm{d}s^2
\equiv g_{\mu\nu}\mathrm{d}x^\mu\mathrm{d}x^\nu
=a^2(\eta)\left(-N^2\mathrm{d}\eta^2+\mathrm{d}\Omega_3^2\right)
\equiv-a^2(\eta)N^2\mathrm{d}\eta^2+a^2(\eta)\gamma_{ij}\mathrm{d}y^i\mathrm{d}y^j.
\end{align}
Here, $\eta$ is the conformal time, $N$ is the lapse parameter and $\mathrm{d}\Omega_3^2=\mathrm{d}\psi^2+\sin^2\psi(\mathrm{d}\theta^2+\sin^2\theta\mathrm{d}\varphi^2)$ is the line element of a unit 3-sphere. We decompose the scalar field on the 3-sphere as
\begin{align}
\phi(\eta,y^i)=\sum_{n=1}^\infty\phi_n(\eta)Q_n(y^i)&=\frac{1}{a(\eta)}\sum_{n=1}^\infty\chi_n(\eta)Q_n(y^i),\\
\int\mathrm{d}\Omega_3Q_nQ_{n'}^*=\delta_{nn'},&\quad \gamma^{ij}\nabla_i\nabla_jQ_n=-(n^2-1)Q_n.
\end{align}
in terms of the spherical harmonics $Q_{nlm}(y^i)$ on 3-sphere, where $y^i$ are the three spherical angles and the indices $l, m$ are suppressed for brevity. With this decomposition scalar field, one has
\begin{align}
\int\mathrm{d}\Omega_3\,\gamma^{ij}\nabla_i\phi\nabla_j\phi=-\int\mathrm{d}\Omega_3\,\phi\, \gamma^{ij}\nabla_i\nabla_j\phi=\sum_n(n^2-1)\phi_n^2.
\end{align}

\subsection{Wheeler-DeWitt equation}\label{subsec:WDW}

To derive the WDW equation for this model, we start with the total bulk action
\begin{align}
S_\mathrm{bulk}=\int\mathrm{d}^4x\sqrt{-g_4}\left[\frac{R}{2}-3H^2-\frac12(\nabla\phi)^2-\frac12m^2\phi^2-\frac12\xi R\phi^2\right],
\end{align}
After substituting the harmonics expansion for $\phi$, this becomes
\begin{align}\label{eq:Sbulk}
S_\mathrm{bulk}&=\int\mathrm{d}\eta\left(-\frac{6\pi^2}{N}\dot{a}^2+6\pi^2NV\right)+\int\mathrm{d}\eta\frac{\mathrm{d}}{\mathrm{d}\eta}\left(\frac{6\pi^2}{N}a\dot{a}-\sum_n\frac{3\xi}{N}\frac{\dot{a}}{a}\chi_n^2\right)\nonumber\\
&+\int\mathrm{d}\eta\sum_n\left(\frac{\dot{\chi}_n^2}{2N}-\frac{1-6\xi}{N}\frac{\dot{a}}{a}\chi_n\dot{\chi}_n+\frac{1-6\xi}{2N}\frac{\dot{a}^2}{a^2}\chi_n^2-\frac12N\omega_n^2\chi_n^2\right).
\end{align}
Here, the overdot denotes a derivative with respect to conformal time $\eta$, and we have used
\begin{align}\label{eq:Ricci}
R=\frac{6}{a^2}\left(1+\frac{\ddot{a}}{N^2a}\right),
\end{align}
and defined 
\begin{align}
V(a)&=a^2-H^2a^4;\\
\omega_n^2&=n^2+m^2a^2+6\xi-1.
\end{align}
The second term in \eqref{eq:Sbulk} cancels out the Gibbons-Hawking-York boundary term
\begin{align}\label{eq:Sbdy}
S_\mathrm{bdy}=-\int\mathrm{d}^3y\sqrt{-g_3}(1-\xi\phi^2)K.
\end{align}
The remaining action is $S_\mathrm{tot}=S_\mathrm{bulk}+S_\mathrm{bdy}=\int\mathrm{d}\eta L$, where $L$ is the Lagrangian. 

To canonically quantize the system, it is convenient to introduce a new field $y_n=\chi_na^{6\xi-1}$; then the canonical momenta corresponding to $a$ and $y_n$ are
\begin{align}
P_a&=\frac{\partial L}{\partial\dot{a}}=-\frac{12\pi^2}{N}\dot{a}+\sum_n\frac{6\xi(1-6\xi)}{Na^{12\xi}}\dot{a}y_n^2;\\
P_{y_n}&=\frac{\partial L}{\partial\dot{y}_n}=\frac{a^{2(1-6\xi)}}{N}\dot{y}_n.
\end{align}
After expressing $\dot{a}$ and $\dot{y}_n$ in terms of $P_a$ and $P_{y_n}$, the Hamiltonian ${\cal H}=(\dot{a}P_a+\sum\limits_n\dot{y}_nP_{y_n}-L)/N$ reads 
\begin{align}\label{eq:Hamiltonian}
{\cal H}=&-\frac{P_a^2}{24\pi^2-12\xi(1-6\xi)a^{-12\xi}y_n^2}-6\pi^2V(a)+\frac{P_{y_n}^2}{2a^{2(1-6\xi)}}+\frac12\omega_n^2a^{2(1-6\xi)}y_n^2
\end{align}
Here the sum over $n$ in each term is left implicit for brevity. The WDW equation 
\begin{align}\label{eq:WDW} 
{\cal H}\Psi(a,y_n)=0
\end{align}
is then obtained by replacing  $P_a\rightarrow-i\hbar\frac{\partial}{\partial a}$ and $P_{y_n}\rightarrow-i\hbar\frac{\partial}{\partial y_n}$. 

We treat the scale factor $a$ as a semiclassical variable and the mode amplitudes $y_n$ as small perturbations.  The WDW equation can then be solved with the following ansatz \cite{Halliwell:1984eu,Wada:1986uy,Vachaspati:1988as,Hong:2002yf}
\begin{align}\label{eq:WKB}
\Psi(a,y_n)=A\exp\left[-\frac{12\pi^2}{\hbar}S(a)-\frac{1}{2\hbar}\sum_nR_n(a)a^{2(1-6\xi)}y_n^2\right].
\end{align}
Substituting this in \eqref{eq:WDW} and neglecting terms $\mathcal{O}(\hbar)$ and $\mathcal{O}(y_n^4)$, we obtain the following equations for the functions $S(a)$ and $R_n(a)$,
\begin{align}
\left(\frac{\mathrm{d}S}{\mathrm{d}a}\right)^2-V(a)&=0;\label{eq:WDW1}\\
a^2\left(\frac{\mathrm{d}S}{\mathrm{d}a}\right)\left(\frac{\mathrm{d}R_n}{\mathrm{d}a}\right)+2a(1-6\xi)\left(\frac{\mathrm{d}S}{\mathrm{d}a}\right)R_n-a^2R_n^2+6\xi(1-6\xi)V(a)+a^2\omega_n^2&=0.\label{eq:WDW2}
\end{align}

\subsection{Solution of WDW equation}\label{subsec:mode}

The solution of WDW equation depends on one's choice of boundary conditions. We shall first focus on the tunneling wave function; the no-boundary wave function will be discussed in Sec.\ref{subsec:HH}.

In the classically allowed region $a>a_*\equiv1/H$, where $V(a)<0$, the tunneling wave function contains only outgoing waves, which describe expanding universes. The corresponding solution of \eqref{eq:WDW1} for $S(a)$ is
\begin{align}
S(a)=i\int_{a_*}^a\mathrm{d}a'\sqrt{-V(a')}+C,
\end{align}
where $C$ is an integration constant.
In the classically forbidden region with $0<a<a_*$ we have exponentially growing and decreasing solutions,
\begin{align}
S^\pm(a)&=\mp\int_a^{a_*}\mathrm{d}a'\sqrt{V(a')}+C,\\
&=\mp\frac{1}{3H^2}\left(1-H^2a^2\right)^{3/2}+C,\label{eq:Spm} 
\end{align}
where the upper and lower signs correspond to decreasing and growing branches, respectively.
The full under-barrier wave function is a superposition of terms
\begin{align}\label{eq:Psipm}
\Psi_\pm(a, \chi_n)&=A_\pm\exp\left[-\frac{12\pi^2}{\hbar}S^\pm(a)-\frac{1}{2\hbar}\sum_nR_n^\pm(a)\chi_n^2\right].
\end{align}
The solutions \eqref{eq:WKB} and \eqref{eq:Psipm} can be matched at the turning point $a_*$ \cite{Vachaspati:1988as}. One finds that the following relations should be satisfied:
\begin{align}
A=A_+=2iA_-,\quad S(a_*)=S^\pm(a_*),\quad R_n(a_*)=R_n^\pm(a_*).
\end{align}

In the classically forbidden region $a<a_*$, it is convenient to switch to the Euclidean conformal time $\tau$ defined by
\begin{equation}\label{eq:tau}
\frac{\mathrm{d}a}{\mathrm{d}\tau}=\left\{\begin{matrix}
+\sqrt{V(a)} ,& \tau<\tau_*;\\
-\sqrt{V(a)} ,& \tau>\tau_*,
\end{matrix}\right.
\end{equation}
which is related to the Lorentzian conformal time $\eta$ via $i\tau=N\eta$. Solving \eqref{eq:tau} explicitly, one has
\begin{align}
\tau=\mp i\,\mathrm{arccot}\sqrt{-1+a^2H^2}+C, \quad C=\tau_*\mp\frac{i\pi}{2}
\end{align}
where the constant is fixed by $\tau(a_*)=\tau_*$. Therefore, the scale factor is solved as
\begin{align}
a(\tau)=\frac{1}{H\cosh(\tau-\tau_*)},
\end{align}
where one can fix $\tau_*=0$ so that $a(\tau)$ is an even function of $\tau$, namely $a(\tau_-)=a(\tau_+)$ for $\tau_-=-\tau_+>0$. With this convention for the two branches of Euclidean conformal time $\tau_\pm$, one finds
\begin{align}
\frac{\mathrm{d}a}{\mathrm{d}\tau_\pm}=\pm\sqrt{V(a(\tau_\pm))}, \quad \frac{\mathrm{d}S^\pm}{\mathrm{d}a}=\pm\sqrt{V(a)},\quad \frac{\mathrm{d}S^\pm}{\mathrm{d}\tau_\pm}=\frac{\mathrm{d}S^\pm}{\mathrm{d}a}\frac{\mathrm{d}a}{\mathrm{d}\tau_\pm}=V(a(\tau_\pm)),
\end{align}
namely the decreasing wave function $\Psi_+$ with $S^+$ and $R_n^+$ is evaluated on the $\tau_+$ branch, and the increasing wave function $\Psi_-$ with $S^-$ and $R_n^-$ is evaluated on the $\tau_-$ branch. 

With these definitions we have
\begin{align}
\left(\frac{\mathrm{d}S^\pm}{\mathrm{d}a}\right)\left(\frac{\mathrm{d}R_n^\pm}{\mathrm{d}a}\right)=\left(\frac{\mathrm{d}S^\pm}{\mathrm{d}\tau_\pm}\right)\left(\frac{\mathrm{d}\tau_\pm}{\mathrm{d}a}\right)^2\left(\frac{\mathrm{d}R_n^\pm}{\mathrm{d}\tau_\pm}\right)=\frac{\mathrm{d}R_n^\pm}{\mathrm{d}\tau_\pm},
\end{align}
and Eq.\eqref{eq:WDW2} for $R_n^\pm$ takes the form
\begin{align}
a^2\left(\frac{\mathrm{d}R_n^\pm}{\mathrm{d}\tau_\pm}\right)+2a(1-6\xi)\left(\frac{\mathrm{d}S^\pm}{\mathrm{d}a}\right)R_n^\pm-a^2(R_n^\pm)^2+6\xi(1-6\xi)V(a)+a^2\omega_n^2=0.
\end{align}
This is a Riccati equation. With the ansatz
\begin{align}
R_n^\pm=-\frac{i}{N}\frac{\dot{u}_n}{u_n}; \quad u_n=\nu_na^{6\xi-1},
\end{align}
or
\begin{align}\label{eq:Rnpm}
R_n^\pm=-\frac{1}{\nu_n}\frac{\mathrm{d}\nu_n}{\mathrm{d}\tau_\pm}+\frac{1-6\xi}{a}\frac{\mathrm{d}a}{\mathrm{d}\tau_\pm},
\end{align}
it reduces to a linear equation for $\nu_n$
\begin{align}
\frac{\mathrm{d}^2\nu_n}{\mathrm{d}\tau_\pm^2}=\left(\omega_n^2+\frac{1-6\xi}{2a}\frac{\mathrm{d}V}{\mathrm{d}a}\right)\nu_n,
\end{align}
or explicitly
\begin{align}\label{eq:modeeq}
\frac{\mathrm{d}^2\nu_n}{\mathrm{d}\tau_\pm^2}=\left[n^2+\left(m^2-2(1-6\xi)H^2\right)a^2\right]\nu_n\equiv\Omega_n^2\nu_n.
\end{align}
We recognize that this is the equation for scalar field modes in de Sitter space.

\section{Constraints on the non-minimal coupling}\label{sec:constraint}

\subsection{Regularity condition at small $a$}\label{subsec:inibdy}

For a physically reasonable quantum state, the functions $R_n(a)$ and $R_n^\pm(a)$ should satisfy the regularity conditions
\begin{align}
\mathrm{Re}R_n(a)>0,\quad \mathrm{Re}R_n^\pm(a)>0     
\end{align}
for all values of $a$ and $n$. Otherwise the probability of quantum fluctuations of the mode amplitudes $\chi_n$ would grow with their magnitude and the state would be unstable.

We first check the regularity condition in the limit $a\rightarrow 0$ (or $\tau_\pm\rightarrow\mp\infty$). The general solution of the mode equation \eqref{eq:modeeq} is
\begin{align}\label{eq:generalsolution}
\nu_n(\tau)=A_nP_\lambda^n(z(\tau))+B_nQ_\lambda^n(z(\tau)),
\end{align}
where
\begin{align}
z(\tau)=\tanh\tau, \quad \lambda=-\frac12+\frac12\sqrt{T}, \quad T=1-4\frac{m^2}{H^2}+8(1-6\xi),
\label{zT}
\end{align}
and $P_\lambda^n(z)$, $Q_\lambda^n(z)$ are the associated Legendre polynomials of first and second kinds, respectively. Note that the quantity $T$  in Eq.~(\ref{zT}) can be expressed as
\begin{align}
T=9-4\mu^2,
\end{align}
where 
\begin{align}
\mu^2=\frac{m^2+\xi R} {H^2} =\frac{m_{\rm eff}^2}{H^2},
\end{align}
$R=12 H^2$ is the scalar curvature of de Sitter space, and $m_{\rm eff}^2\equiv m^2+\xi R$ is the effective mass squared of the field.
We shall refer to $\mu^2$ as the effective mass parameter.

The asymptotic expansions of $P_\lambda^n(z)$ and $Q_\lambda^n(z)$ at $z\rightarrow\pm1$ are
\begin{align}
P_\lambda^n(z\sim+1)&\sim-\frac{1}{\pi n!}2^{-\frac{n}{2}}\Gamma(n-\lambda)\Gamma(n+\lambda+1)\sin(\lambda\pi)(1-z)^{\frac{n}{2}}[1+\mathcal{O}(1-z)];\label{eq:Pzp1}\\
Q_\lambda^n(z\sim+1)&\sim(-1)^n2^{\frac{n}{2}-1}(n-1)!(1-z)^{-\frac{n}{2}}[1+\mathcal{O}(1-z)];\label{eq:Qzp1}\\
P_\lambda^n(z\to-1)&-\frac{(n-1)!}{\pi}2^{\frac{n}{2}}\sin(\lambda\pi)(1+z)^{-\frac{n}{2}}[1+\mathcal{O}(1+z)];\label{eq:Pzm1}\\
Q_\lambda^n(z\sim-1)&\sim-2^{\frac{n}{2}-1}(n-1)!\cos(\lambda\pi)(1+z)^{-\frac{n}{2}}[1+\mathcal{O}(1+z)],\label{eq:Qzm1}
\end{align}
where $\psi(z)\equiv\Gamma'(z)/\Gamma(z)$ is the digamma function and $\gamma$ is the Euler constant. 
Then it is easy to see that
\begin{align}
\left.\frac{1}{\nu_n}\frac{\mathrm{d}\nu_n}{\mathrm{d}\tau}\right|_{\tau\rightarrow+\infty}&\sim-\frac{n}{2}\frac{\mathrm{sech}^2\tau}{1-\tanh\tau}\frac{A_nP_\lambda^n-B_nQ_\lambda^n}{A_nP_\lambda^n+B_nQ_\lambda^n}=-n\frac{A_nP_\lambda^n-B_nQ_\lambda^n}{A_nP_\lambda^n+B_nQ_\lambda^n};\\
\left.\frac{1}{\nu_n}\frac{\mathrm{d}\nu_n}{\mathrm{d}\tau}\right|_{\tau\rightarrow-\infty}&\sim-\frac{n}{2}\frac{\mathrm{sech}^2\tau}{1+\tanh\tau}\frac{A_nP_\lambda^n+B_nQ_\lambda^n}{A_nP_\lambda^n+B_nQ_\lambda^n}=-n,
\end{align}
and
\begin{align}
R_n^+(z\to-1)&=-\left.\frac{1}{\nu_n}\frac{\mathrm{d}\nu_n}{\mathrm{d}\tau}\right|_{\tau\rightarrow-\infty}+(1-6\xi)=n+(1-6\xi);\\
R_n^-(z\to+1)&=-\left.\frac{1}{\nu_n}\frac{\mathrm{d}\nu_n}{\mathrm{d}\tau}\right|_{\tau\rightarrow+\infty}-(1-6\xi)=n\frac{A_nP_\lambda^n(z\to+1)-B_nQ_\lambda^n(z\to+1)}{A_nP_\lambda^n(z\to+1)+B_nQ_\lambda^n(z\to+1)}-(1-6\xi).
\end{align}
$Q_\lambda^n(z\to+1)$ dominates over $P_\lambda^n(z\to+1)$, so for $B_n\neq0$ one finds $R_n^-(z\to+1)=-n-(1-6\xi)=-R_n^+(z\to-1)$.  Therefore, $R_n^-(z\to+1)$ and $R_n^+(z\to-1)$ cannot be simultaneously non-negative.
Hence, one has to choose $B_n=0$, and the final solution is
\begin{align}\label{eq:1stsol}
\nu_n(\tau)=P_\lambda^n(\tanh\tau). 
\end{align}
Note that this choice of mode functions corresponds to the standard Bunch-Davies vacuum. 
With this choice, Eq.~\eqref{eq:Rnpm} gives
\begin{align}\label{eq:Rnpma0}
R_n^\pm (a\to 0)= n \pm (1-6\xi),
\end{align}
and it is easily seen that in order for the regularity condition to be satisfied for all $n$, the curvature coupling parameter should be in the range
\begin{align}\label{eq:xibounda0}
0\leq\xi\leq\frac13.
\end{align}
Note that the previously studied cases of minimal coupling $\xi=0$ and conformal coupling $\xi=\frac16$ satisfy these bounds. Note also that the constraints \eqref{eq:xibounda0} come from the $n=1$ mode. Higher modes would give weaker constraints.

The following caveat should be noted in the above analysis.  The leading order terms in $(1+z)$ in Eqs.~\eqref{eq:Pzm1} and \eqref{eq:Qzm1} have the same power $-n/2$, so one can construct a combination
\begin{align}\label{eq:2ndsol}
X_\lambda^n(z)=P_\lambda^n(z)-\frac{2}{\pi}\tan(\lambda\pi)Q_\lambda^n(z)
\end{align}
where the leading terms cancel out.  
If $X_\lambda^n(z)$ are chosen as the mode functions $\nu_n(z)$, we have verified numerically that
\begin{align}
\left.\frac{1}{\nu_n}\frac{d\nu_n}{d\tau}\right|_{z\to\pm 1} = n,
\end{align}
and the regularity conditions at $z\to\pm 1$ are 
\begin{align}
{\tilde R}_n(z\to\pm 1)=-n\mp (1-6\xi) \geq 0, 
\label{z-1}
\end{align}
where tilde indicates that ${\tilde R}_n(z)$ are calculated using the mode functions $X_\lambda^n(z)$.
Clearly, these two conditions cannot be simultaneously satisfied, and it appears that $X_\lambda^n(z)$ is not an acceptable choice of mode functions.  We will see however (in Sec.~\ref{subsec:subdominant}) that for some parameter values it is inconsistent to keep the subdominant growing branch of the wave function, so we cannot impose the regularity condition at $z\to +1$.  The remaining regularity condition (\ref{z-1}) at $z\to -1$ requires that $n\leq 1-6\xi$.  It can be satisfied for $\xi\leq 0$ and for sufficiently small values of $n$.  This may allow one to construct regular quantum states in models where the canonical Bunch-Davies state (\ref{eq:1stsol}) violates the regularity conditions.  We shall discuss an example of this in Sec.~\ref{subsec:subdominant}.

\subsection{Regularity condition at the turning point}\label{subsec:turnpt}

\begin{figure}
\centering
\includegraphics[width=0.49\textwidth]{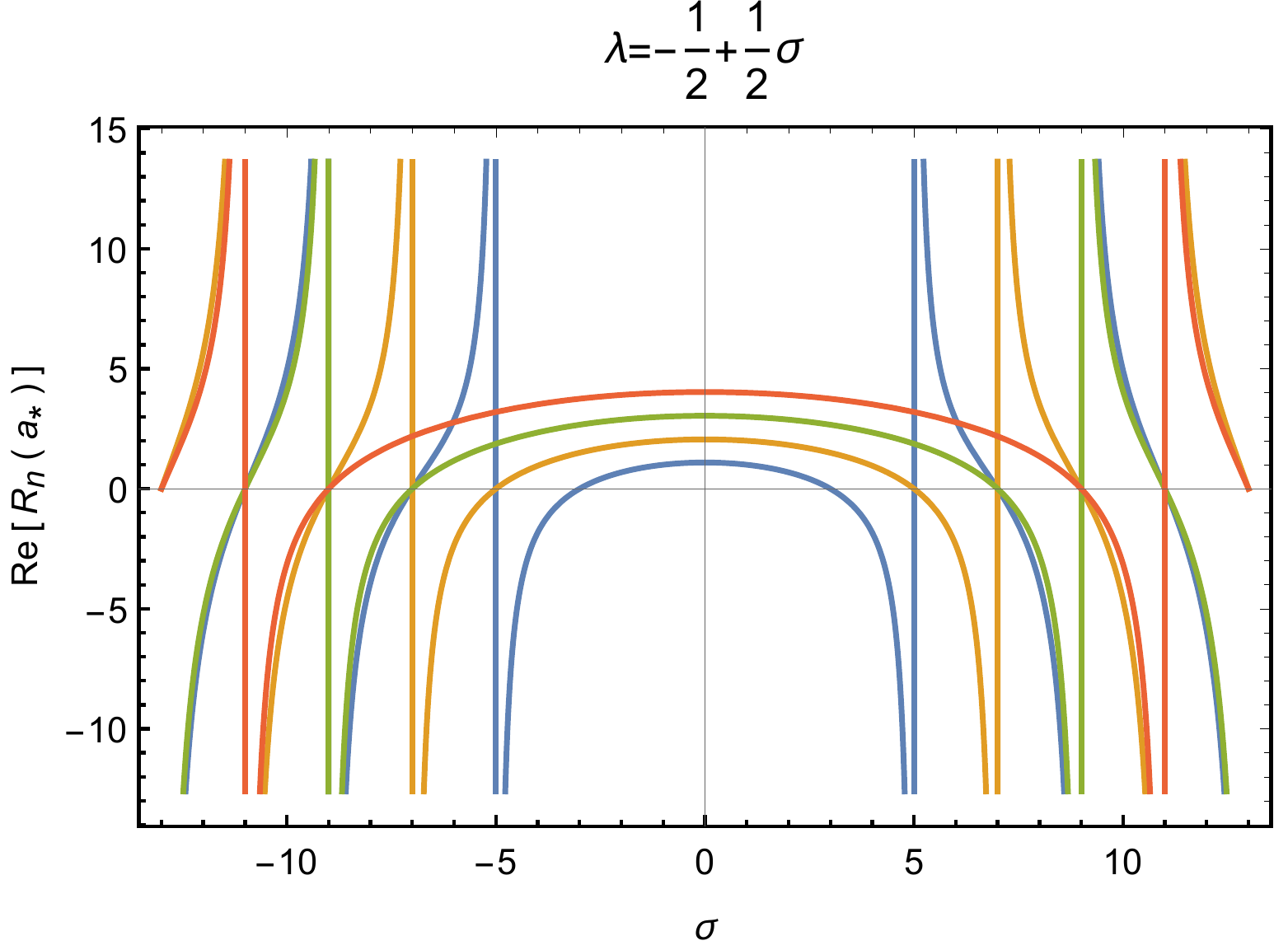}
\includegraphics[width=0.49\textwidth]{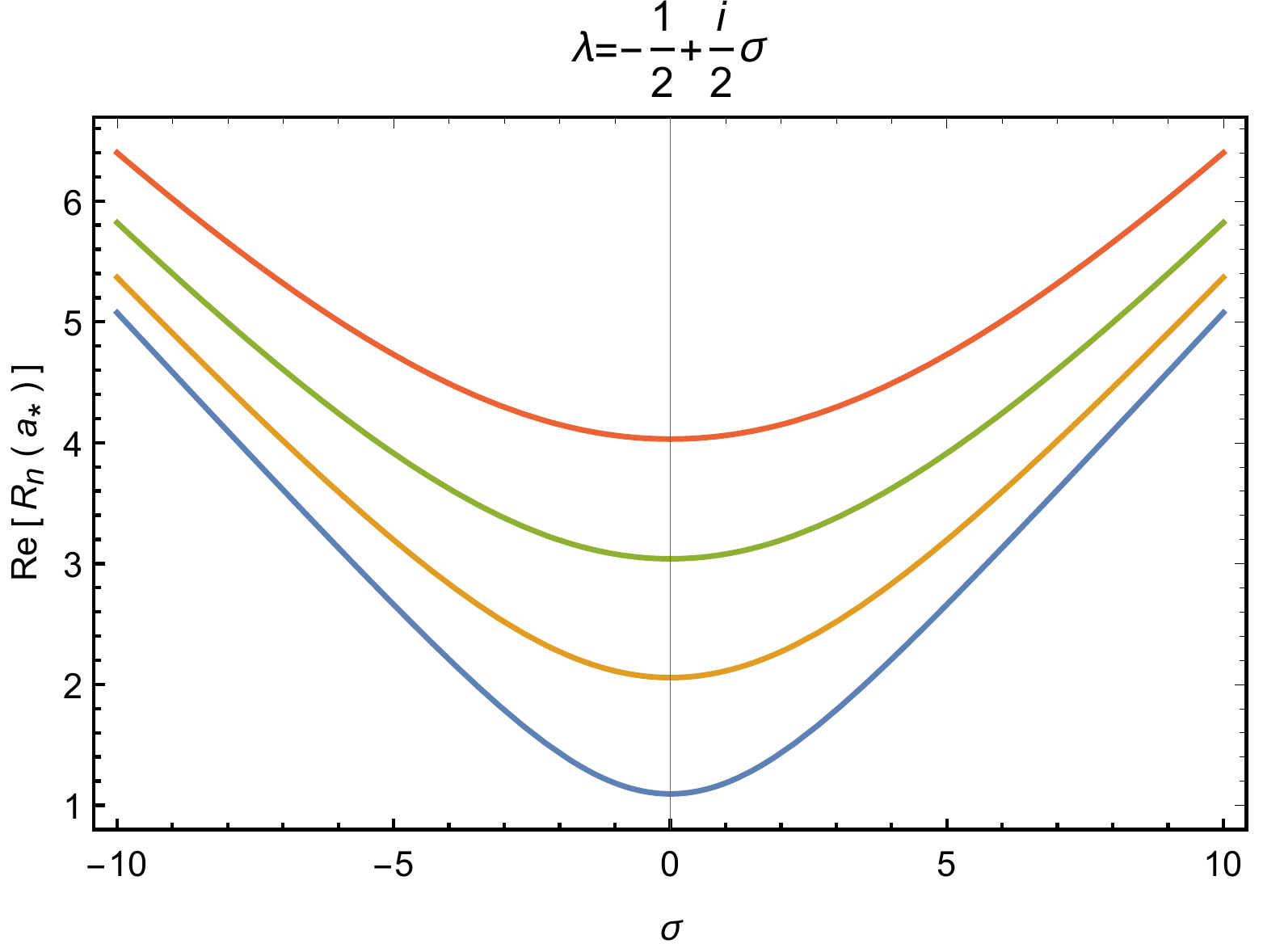}\\
\caption{The behaviour of $\mathrm{Re}R_n(a_*)$ with respect to $\sigma$, where $\sigma$ is defined by $\lambda=-\frac12+\frac12\sigma$ when $T\geq0$ (left), and by $\lambda=-\frac12+\frac{i}{2}\sigma$ when $T<0$ (right). The $n=1, 2, 3, 4$ modes are shown in blue, yellow, green and red, respectively.}\label{fig:Tpm}
\end{figure}

We next check the regularity condition at the turning point $a=a_*\equiv1/H$ (that is, $\tau=0$ and $z=\tanh\tau=0$). With the mode function solutions \eqref{eq:1stsol}, one immediately derives
\begin{align}
R_n(a_*)
=-\left.\frac{1}{\nu_n}\frac{\mathrm{d}\nu_n}{\mathrm{d}\tau}\right|_{\tau\rightarrow0}
=\left(\lambda z-(\lambda+n)\frac{P_{\lambda-1}^n(z)}{P_\lambda^n(z)}\right)_{z\rightarrow0}
=-(\lambda+n)\frac{P_{\lambda-1}^n(0)}{P_\lambda^n(0)}.
\end{align}
There are two cases to consider:
\begin{align}
T\geq0 :&\quad\lambda=-\frac12+\frac12\sigma,\\
T<0 :&\quad\lambda=-\frac12+\frac{i}{2}\sigma,
\end{align}
where $\sigma$ is real, which are illustrated in Fig.\ref{fig:Tpm} for a few representative low-$n$ modes. 
If $T<0$, the regularity condition is satisfied for all $n$-modes, as indicated in the right panel of Fig.\ref{fig:Tpm}. This can also be seen from the explicit formula
\begin{align}
\mathrm{Re}R_n(a_*)=2\frac{\left|\Gamma\left(\frac34-\frac{n}{2}+\frac{i}{4}\sigma\right)\right|^2}{\left|\Gamma\left(\frac14-\frac{n}{2}+\frac{i}{4}\sigma\right)\right|^2}\geq0,
\end{align}
where we have used 
\begin{align}
P_\lambda^n(0)=\frac{2^n\sqrt{\pi}}{\Gamma\left(\frac{1-n-\lambda}{2}\right)\Gamma\left(1-\frac{n-\lambda}{2}\right)}.
\end{align}

On the other hand, if $T\geq0$, there is a violation of regularity, $R_n(a_*)<0$, for $(2n+1+2k)<|\sigma|<(2n+3+2k)$ with $k=0,2,4,6\cdots$ as indicated in the left panel of Fig.\ref{fig:Tpm}. Therefore, the regularity condition is satisfied for all $n$ if $|\sigma|<\sigma_0$, where $\sigma_0 =3$ is the smallest positive solution of
\begin{align}
\mathrm{Re}R_{n=1}(a_*)=-(\lambda+1)\frac{P_{\lambda-1}^1(0)}{P_\lambda^1(0)}=\lambda\frac{P_{\lambda+1}^1(0)}{P_\lambda^1(0)}=0.
\end{align} 

Thus the constraint on the curvature coupling that comes from the regularity condition at the turning point is $T\leq\sigma_0^2=9$, or
\begin{align}\label{eq:tachyon}
\mu^2 \geq0.
\end{align} 
This condition requires that the effective mass of the field is $m_\mathrm{eff}^2 \geq 0$, thus avoiding a tachyonic instability.

It has been shown in the case of minimally and conformally coupled fields that the sign of the functions ${\rm Re} R_n(a)$ remains the same in the entire classically allowed region \cite{Hong:2002yf,Damour:2019iyi}. The same proof goes through for the general coupling $\xi$.  We conclude that the regularity condition holds in the classically allowed region if and only if the coupling $\xi$ lies in the range given by \eqref{eq:tachyon}.

\subsection{Regularity condition in the whole tunneling region}\label{subsec:forbidden}

\begin{figure}
\centering
\includegraphics[width=0.49\textwidth]{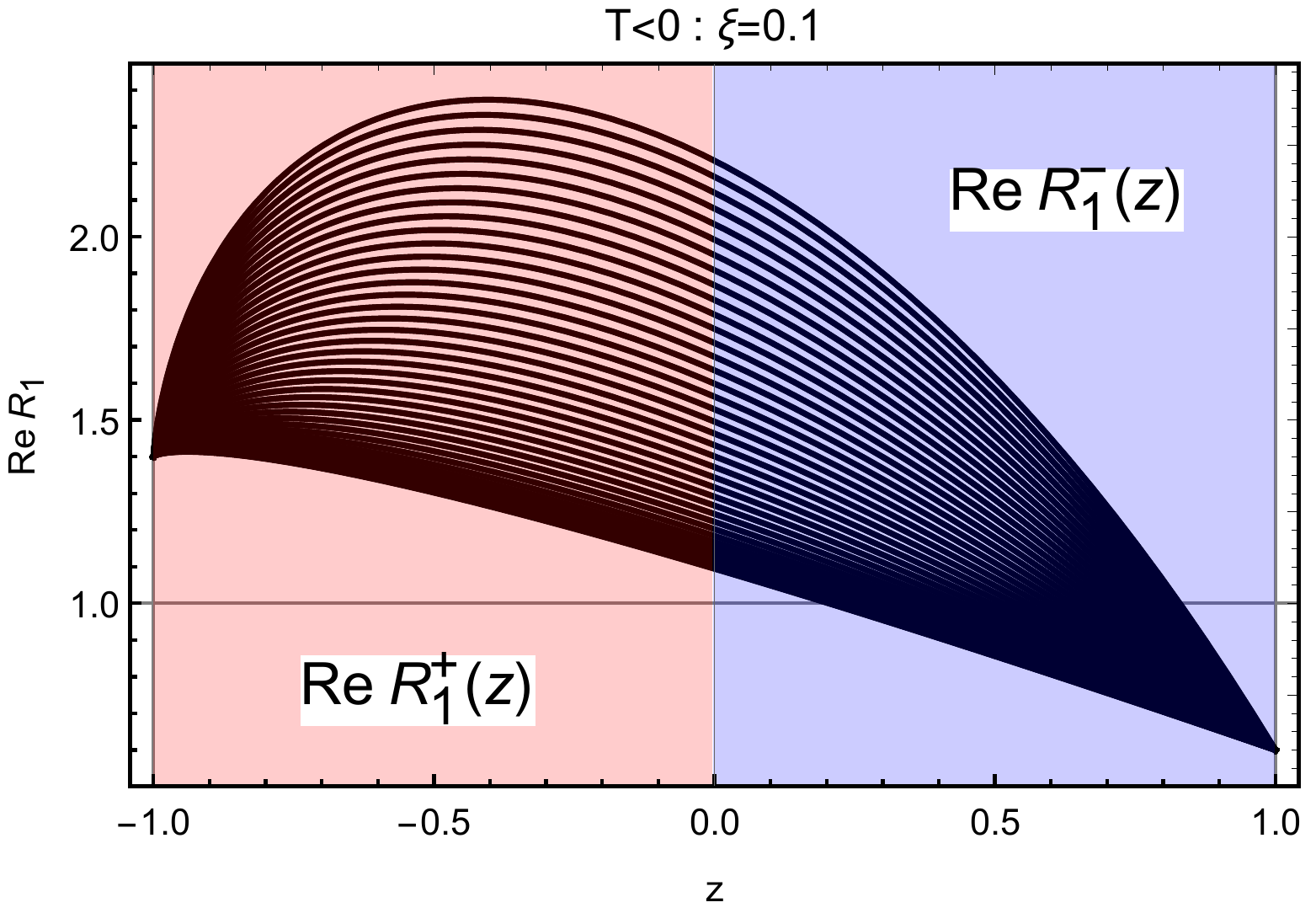}
\includegraphics[width=0.49\textwidth]{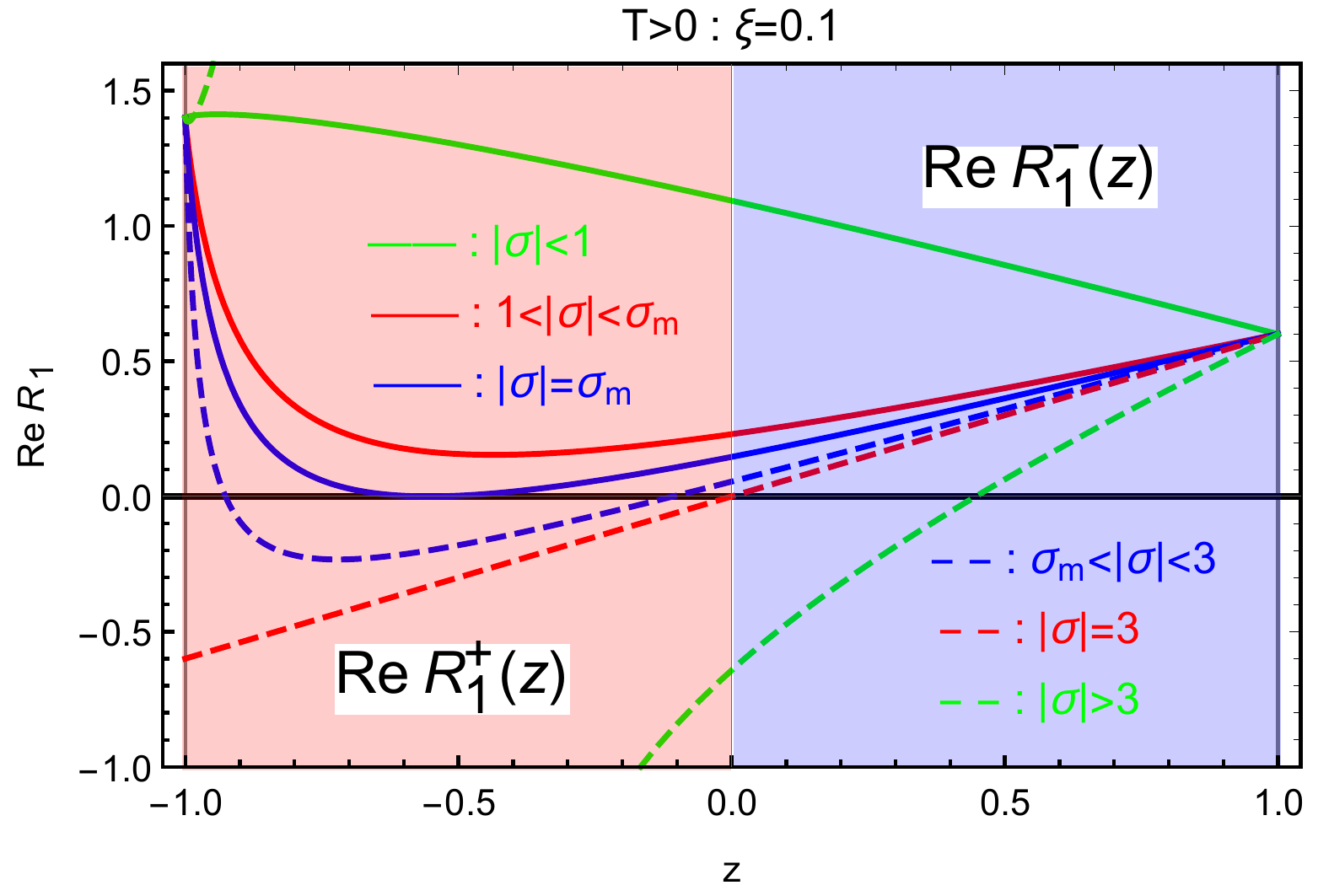}\\
\includegraphics[width=0.49\textwidth]{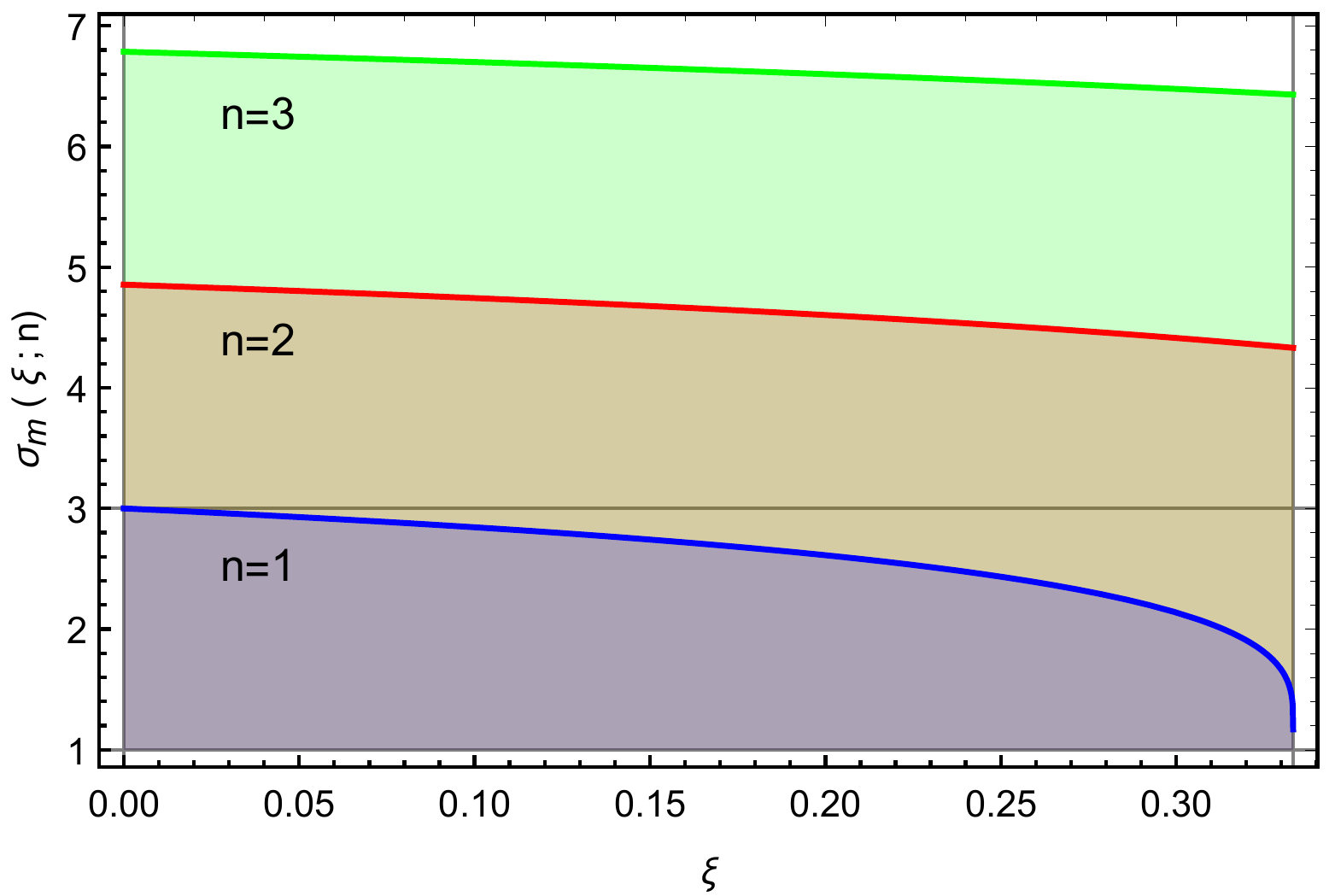}
\includegraphics[width=0.49\textwidth]{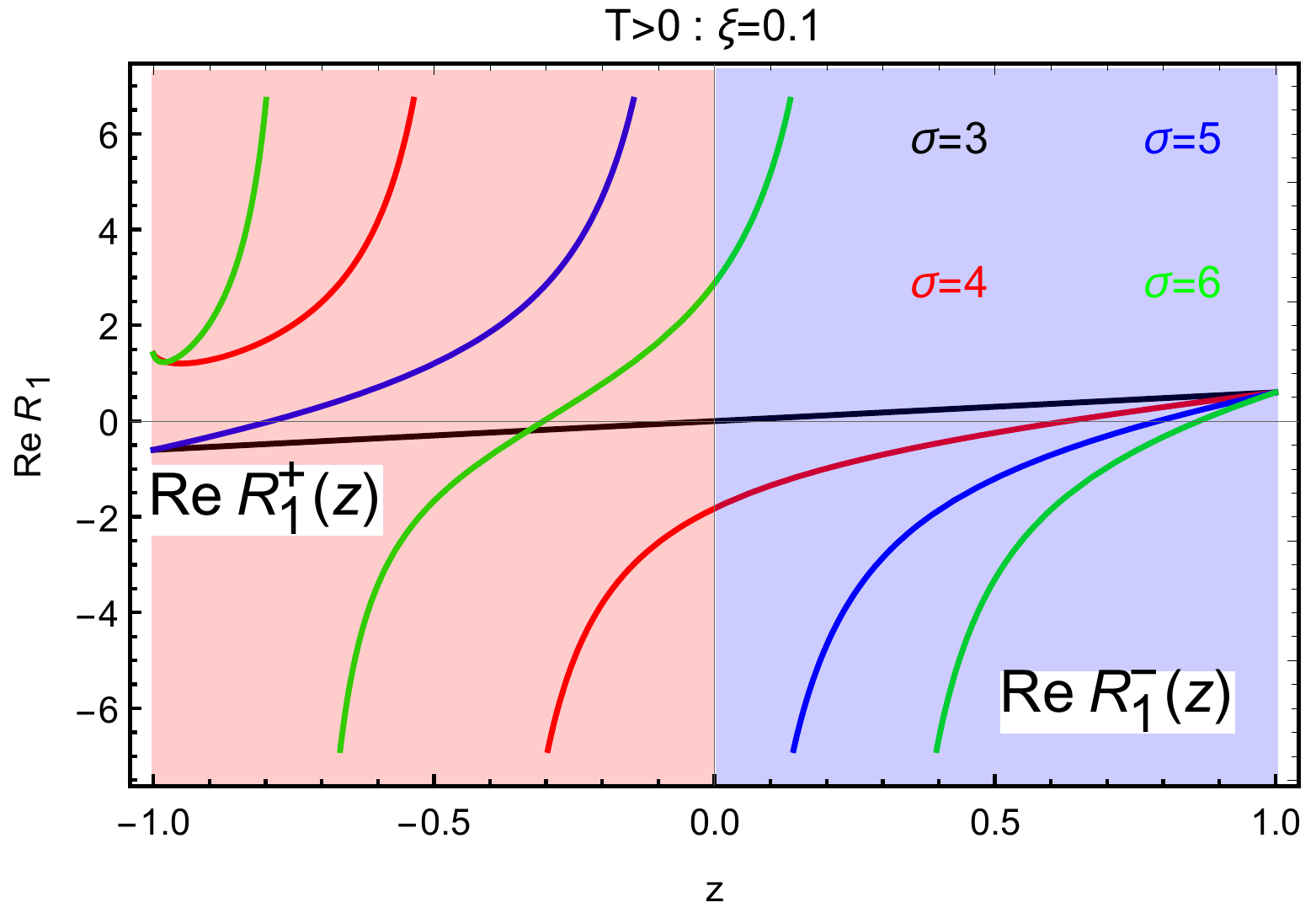}\\
\caption{In the upper two panels we show the functions $\mathrm{Re}R_n(z)$ for the mode $n=1$ and the coupling parameter $\xi=0.1$.  The ranges $z<0$ and $z>0$ correspond to the functions $\mathrm{Re}R_n^+(z)$ (red shaded region) and $\mathrm{Re}R_n^-(z)$ (blue shaded region), respectively.  The left panel represents the case of $T<0$ with $-4\leq\sigma\leq4$, while the right panel represents the case of $T>0$ with several values of $|\sigma|$ below and above the critical value $\sigma_m$. In the right panel we also show some plots of $\mathrm{Re}R_n(z)$ with $|\sigma|$ at, above and below the value of $|\sigma|=3$. In the lower left panel we show the critical value $\sigma_m$ as a function of $\xi$ for some representative modes $n=1,2,3$. In the last panel, we illustrate the divergent behaviour for some parameter choices. For $\sigma_m<|\sigma|\leq3$, only $\mathrm{Re}R_1^+$ violates the regularity condition. For $3<|\sigma|<5$, $\mathrm{Re}R_1^+$ is divergent at some intermediate $z$ while $\mathrm{Re}R_1^-$ violates the regularity condition. When $|\sigma|=5$, both $\mathrm{Re}R_1^+$ and $\mathrm{Re}R_1^-$ are divergent at $z=0$. For $|\sigma|>5$, both $\mathrm{Re}R_1^+$ and $\mathrm{Re}R_1^-$ are divergent and negative at some intermediate $z$.} 
\label{fig:Rnofa}
\end{figure}

For now, we have only checked the regularity condition at $a=0$ and $a\geq a_*$; hence
the bounds \eqref{eq:xibounda0} and \eqref{eq:tachyon} on the curvature coupling parameter only serve as necessary conditions for the regularity of the wave function. 

To find sufficient conditions, we need to examine the behavior of
\begin{align}
R_n^\pm(z_\pm=\tanh\tau_\pm)&=-\frac{1}{\nu_n}\frac{\mathrm{d}\nu_n}{\mathrm{d}\tau_\pm}+\frac{1-6\xi}{a}\frac{\mathrm{d}a}{\mathrm{d}\tau_\pm};\nonumber\\
&=\lambda z_\pm-(\lambda+n)\frac{P_{\lambda-1}^n(z_\pm)}{P_\lambda^n(z_\pm)}-(1-6\xi)z_\pm
\label{eq:Rnz}
\end{align}
in the whole classically forbidden range $0<a<a_*$ \footnote{Note that our conclusions about the behavior of $R_n^\pm$ as functions of $\sigma$ are based on sampling different characteristic values of $\sigma$ without a general proof.}.
For $T<0$, or $\lambda=-\frac12+\frac{i}{2}\sigma$ with $\sigma$ real, we have verified, by sampling a large number of parameter values, that $\mathrm{Re}R_n^\pm(z)$ is a concave function. This is illustrated in the first panel of Fig.\ref{fig:Rnofa} for the mode $n=1$ with $\xi=0.1$ and $-4\leq\sigma\leq4$. In this case, the regularity condition $\mathrm{Re}R_n^\pm(z)\geq0$ is satisfied in the whole classically forbidden region, as long as it is satisfied at both of its boundaries. 

\begin{figure}
\centering
\includegraphics[width=0.8\textwidth]{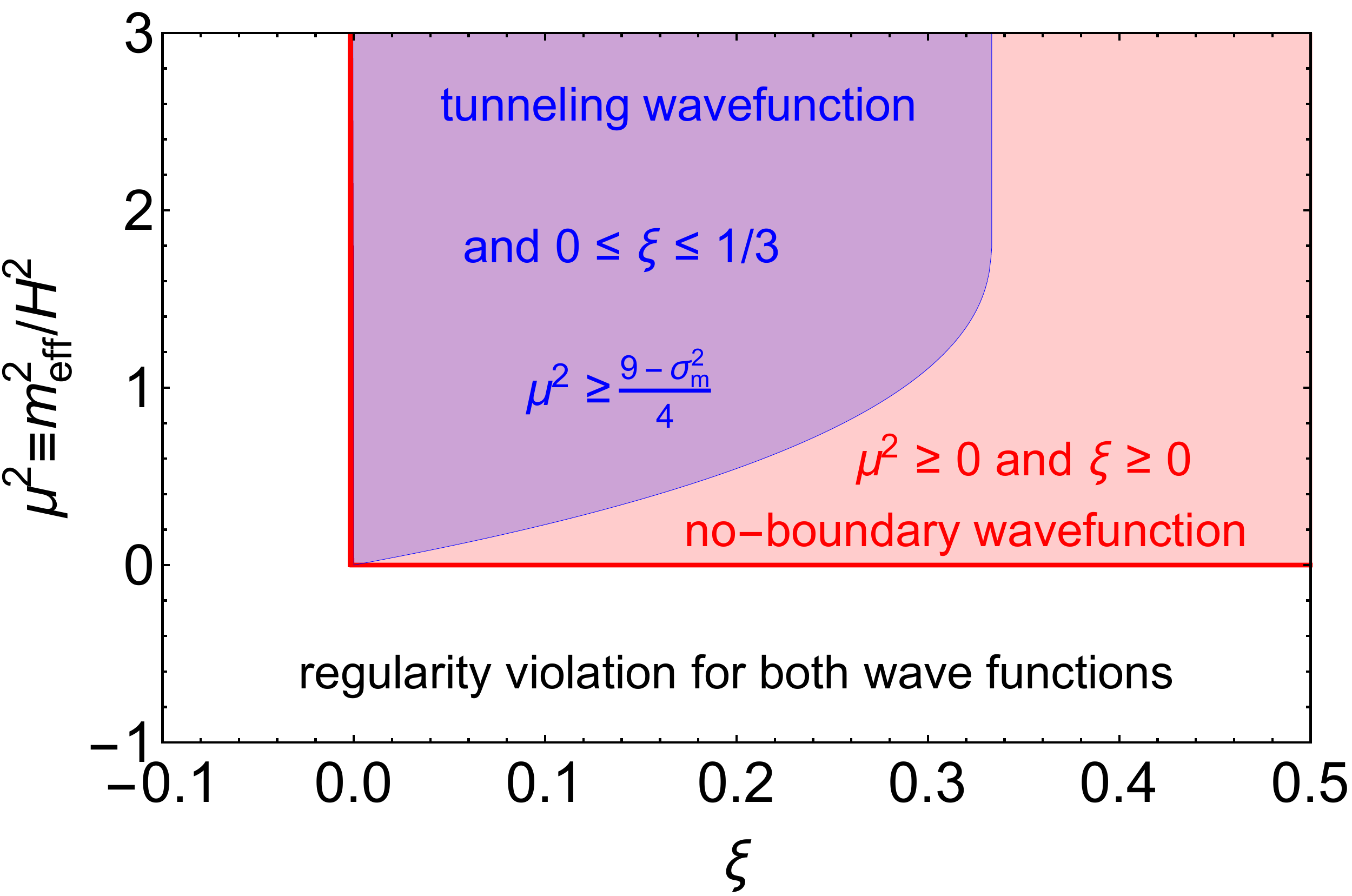}\\
\caption{Parameter values satisfying the necessary and sufficient conditions of regularity for the tunneling wave function are shown by blue shading. 
Those for the no-boundary wave function are shown by blue and red shading. 
In the white region the regularity conditions are violated for both wave functions. 
}
\label{fig:xibound}
\end{figure}

For $T>0$, or $\lambda=-\frac12+\frac12\sigma$, there may be a critical value $\sigma_m$ for the absolute value of $\sigma$, above which the regularity condition is violated in some range of $z$. This is illustrated in the second panel of Fig.\ref{fig:Rnofa} for the mode $n=1$ and the coupling parameter $\xi=0.1$. We note that the $z<0$ and $z>0$ parts of the graphs in this panel correspond to $R_n^+(z)$ and $R_n^-(z)$ functions, respectively. It is clear from the figure that the critical value $\sigma_m$ occurs when
\begin{align}
R_n^\pm(z_m)=0,\quad \left.\frac{\mathrm{d}R_n^\pm}{\mathrm{d}z}\right|_{z=z_m}=0.
\end{align}
Note that $R_n^\pm (z)$ are real in this case if both $\lambda$ and $n$ are real. 
Substituting $R_n^\pm (z)$ from Eq.~\eqref{eq:Rnz} we obtain
\begin{align}
\lambda z_m-(\lambda+n)\frac{P_{\lambda-1}^n(z_m)}{P_\lambda^n(z_m)}-(1-6\xi)z_m&=0;\\
\lambda-\frac{\lambda+n}{z_m^2-1}\left[(\lambda-n)-2\lambda z_m\frac{P_{\lambda-1}^n(z_m)}{P_\lambda^n(z_m)}+(\lambda+n)\left(\frac{P_{\lambda-1}^n(z_m)}{P_\lambda^n(z_m)}\right)^2\right]-(1-6\xi)&=0.
\end{align}
The critical value $\sigma_m(\xi ; n)$ can now be found from
\begin{align}\label{eq:zm}
\frac{P_{-\frac32+\frac12\sigma_m}^n(z_m)}{P_{-\frac12+\frac12\sigma_m}^n(z_m)}=\frac{\lambda-(1-6\xi)}{n+\lambda}z_m,\quad z_m=-\sqrt{\frac{4n^2+5-24\xi-\sigma_m^2}{(3-12\xi)^2-\sigma_m^2}},
\end{align}
as a function of mode number $n$ and coupling parameter $\xi$. It is clear from the third panel of Fig.\ref{fig:Rnofa} that in order for the regularity condition to hold in the entire under-barrier region, it is necessary and sufficient for $\sigma$ to satisfy $|\sigma|\leq\sigma_m(\xi;1)\equiv\sigma_m$. This bound can be translated into a constraint on the effective mass,
\begin{align}\label{eq:lowerbound0}
\mu^2 \geq(9-\sigma_m^2)/4. 
\end{align}
This condition automatically guarantees the necessary condition \eqref{eq:tachyon} at the turning point (since $\sigma_m\leq3$), which in turn ensures the regularity in the classically allowed region. Summarizing our results, the allowed parameter range for the tunneling wave function is shown as the blue shaded region in Fig.\ref{fig:xibound}.

It is worth noting that for some parameter choices $\mathrm{Re}R_1^\pm$ becomes divergent at some intermediate $z$. 
This is illustrated for some $n=1$ modes in the last panel of Fig.\ref{fig:Rnofa}. Specifically, for $\sigma_m<|\sigma|\leq3$, only $\mathrm{Re}R_1^+$ turns negative in some intermediate range of $z$ while $\mathrm{Re}R_1^-$ stays non-negative with both $\mathrm{Re}R_1^\pm$ remaining finite; For $3<|\sigma|<5$, $\mathrm{Re}R_1^+$ becomes divergent at some intermediate $z$ while $\mathrm{Re}R_1^-$ turns negative in some intermediate range of $z$; For $|\sigma|=5$, both $\mathrm{Re}R_1^+$ and $\mathrm{Re}R_1^-$ are divergent at $z=0$; For $|\sigma|>5$, both $\mathrm{Re}R_1^+$ and $\mathrm{Re}R_1^-$ become divergent and negative at some intermediate $z$. In conclusion, 
$\mathrm{Re}R_1^+$ becomes divergent if
\begin{align}
|\sigma|>3\Rightarrow \mu^2<0.
\label{xigeq0}
\end{align}

\subsection{Discarding the subdominant branch}\label{subsec:subdominant}

The growing branch of the wave function $\Psi_-(a,\chi_n)$ is exponentially suppressed compared to the decreasing branch at $\chi_n=0$.  So keeping this branch while we neglect larger corrections to the WKB formula requires a special justification.  This issue was addressed in Ref.~\cite{Vachaspati:1988as} for the case of $m=\xi=0$ with the following argument.  First note that the growing and decreasing branches have comparable magnitudes at the turning point $a=a_*$.  Furthermore, it was shown in \cite{Vachaspati:1988as} that $R_n^+(a) > R_n^-(a)$ for all $a<a_*$.  This means that the function $\Psi_+(a,\chi_n)$ decreases with $\chi_n$ exponentially faster than $\Psi_-(a,\chi_n)$ and therefore $\Psi_-$ dominates at sufficiently large $\chi_n$. 
We thus have a continuous domain in the $\{a,\chi_n\}$ superspace, ranging from $a_*$ to $a=0$, where $\Psi_-$ is non-negligible compared to $\Psi_+$.  If ${\rm Re} R_n^-$ were to become negative somewhere in this domain, this would indicate a violation of regularity.  We shall now extend this analysis to some values of $m$ and $\xi$ other than $m=\xi=0$.
  
We are justified to keep the subdominant branch if $R_n^+(a) > R_n^-(a)$ for $a<a_*$.  This means in particular that
\begin{align}
\frac{\mathrm{d}}{\mathrm{d}z} {\rm Re} R_n(z=0) <0.
\label{d/dz}
\end{align}
If this condition is violated, then $R_n^-$ becomes greater than $R_n^+$ at $a$ immediately below $a_*$ and the growing branch cannot be kept.  Thus, the condition (\ref{d/dz}) is a necessary condition for keeping the growing branch and its violation is a sufficient condition for discarding that branch.  

We have used the regularity condition on the growing branch to derive the constraint $\xi\geq 0$ and to exclude the white region of the parameter space in the left upper corner of Fig.~\ref{fig:xibound}.  We have verified, however, that the condition (\ref{d/dz}) is satisfied everywhere in that region, so its exclusion is consistent with the condition \eqref{d/dz}. It is possible that even though (\ref{d/dz}) is satisfied, the inequality ${\rm Re}R_n^+(a) > {\rm Re}R_n^-(a)$ is violated at some $a<a_*$.  Some parameter values in the white region may then be allowed.  We have not performed a complete analysis in the whole range of $0<a<a_*$.

As we mentioned in Sec.~\ref{subsec:inibdy}, for some parameter values where the Bunch-Davies (BD) state violates the regularity condition it may be possible to construct regular quantum states using the mode functions $X_\lambda^n(z)$  
(\ref{eq:2ndsol}).  An important example is the inflaton field near the maximum of its potential.  In this case $\xi=0$ and $m^2<0$ with $|m^2|\lesssim H^2$, so the regularity condition for the $n=1$ BD mode is violated at the turning point.  A regular quantum state can be obtained using the BD modes for $n\geq 2$ and 
\begin{align}
\nu_1(z)=X_\lambda^1(z).  
\label{nu1}
\end{align}
It is shown in the Appendix that the condition (\ref{d/dz}) is then violated for the $n=1$ mode, so we should only use the regularity condition (\ref{z-1}) at $z\to -1$. This gives $\xi\leq 0$, which is consistent with $\xi=0$. 
Furthermore, we have verified that the mode (\ref{nu1}) satisfies the regularity condition at the turning point.  This mode is actually the same as was used for the inflaton field in Refs.~\cite{Vilenkin:1986cy,Vilenkin:1987kf}.

\subsection{No-boundary wave function}\label{subsec:HH}

The no-boundary wave function includes only the growing branch under the barrier.  In the classically allowed region it includes an expanding and a contracting branch of equal amplitude, which are complex conjugates of one another. 
The matching conditions at $a=a_*$ are
\begin{align}
S(a_*)=S^*(a_*)=S^-(a_*), \quad R_n(a_*)=R_n^*(a_*)=R_n^-(a_*),
\end{align}
and the regularity condition is
\begin{align}
R_n^-(a)>0.
\end{align}
At $a\to 0$, $R_n^-(a)$ is given by Eq.~\eqref{eq:Rnpma0}, and the regularity condition yields the bound $\xi >0$.  At the turning point $a_*$ we have $R_n^+ = R_n^-$, so the bound on $\xi$ from that point is given by Eq.~\eqref{eq:tachyon}, the same as for the tunneling wave function.

We now consider the bound coming from the under-barrier region $0<a<a_*$. In the case of the tunneling wave function, we found that regularity is violated when the parameter $\sigma$ defined in Sec.~\ref{subsec:forbidden} gets larger than the critical value $\sigma_m$.  This violation first occurs at $z_m <0$ on the branch $R_n^+(z)$, which is absent in the no-boundary wave function (see Eq.~\eqref{eq:zm} and the upper right panel of Fig.~\ref{fig:Rnofa}). As $\sigma$ is increased further, the range of $z$ where regularity is violated gets wider, and eventually this range extends into the region of $z>0$, which corresponds to the branch $R_n^-(z)$.  Thus the critical point at which regularity is violated for the no-boundary wave function is determined by the condition $R_n(z=0)=0$.  The point $z=0$ corresponds to $a=a_*$; hence this condition is the same as we derived in Sec.~\ref{subsec:turnpt}, $\mu^2 \geq 0$.  As before, regularity at the turning point guarantees regularity in the entire classically allowed region.  Hence the complete (necessary and sufficient) regularity constraints for the no-boundary wave function read
\begin{align}
\mu^2 \geq 0; ~~~~ \xi\geq 0.
\end{align}
The resulting allowed parameter range is shown as the red shaded region in Fig.~\ref{fig:xibound}.

It follows from the analysis in Sec.~\ref{subsec:forbidden} that 
$\mathrm{Re}R_1^-$ becomes divergent for the no-boundary wave function if
\begin{align}
|\sigma|\geq5\Rightarrow \mu^2 \leq-4.
\end{align}

\section{Summary and discussion}\label{sec:con}

In a minisuperspace context, the regularity condition for the wave function of the universe requires that fluctuations of quantum fields are suppressed, {\it i.e.}, the probability of fluctuations decreases with their amplitude.  In the present paper we analyzed this condition for a de Sitter minisuperspace with a non-minimally coupled scalar field.  We found that the condition is satisfied only if the mass of the field $m$ and the curvature coupling parameter $\xi$ obey certain bounds.   
For the no-boundary wave function the bounds are
\begin{align}\label{eq:tachyon2}
\mu^2 \geq 0
\end{align}
and
\begin{align}\label{eq:lowerbound}
\xi\geq 0.
\end{align}
For the tunneling wave function the constraints are stronger; the allowed range of $m$ and $\xi$ in this case is shown as the blue shaded region in the right lower panel of Fig.~\ref{fig:Rnofa}. In particular, in addition to the lower bound \eqref{eq:lowerbound}, $\xi$ must also obey an upper bound 
\begin{align}\label{eq:upperbound}
\xi\leq 1/3.
\end{align}
The bound \eqref{eq:tachyon2} has a clear physical interpretation: the effective mass of the scalar field must satisfy $m_\mathrm{eff}^2 =m^2 +\xi R \geq 0$ to avoid tachyonic instability. However, the origin of the other bounds is not clear. Violation of these additional bounds results in a violation of regularity only in the classically forbidden, under-barrier region. It appears, however, that quantum states with unbounded fluctuations are unacceptable even under the barrier.

Our regularity conditions typically select the Bunch-Davies (BD) state of the scalar field, and the above bounds generally apply only for this choice of state.  We found, however, that for some parameter values of the tunneling wave function it is possible to construct
other regular states with different low-$n$ modes and to evade some of the bounds.  One example is an inflaton field near the maximum of its potential, when $\xi=0$ and $m^2<0$.  This case is important because the tunneling wave function is peaked near the maximum of the potential.  For the no-boundary wave function the BD state is always selected and the bounds always apply. 

We finally comment on the intriguing question about the relation between the constraints on non-minimal coupling in Jordan and Einstein frames \footnote{Note that the classical/quantum (in)equivalence between Jordan and Einstein frames is a long-standing issue \cite{Catena:2006bd,Faraoni:2006fx,Chiba:2008ia,Deruelle:2010ht,Nozari:2010uu,Gong:2011qe,George:2013iia,Chiba:2013mha,Jarv:2014hma,Li:2014qwa,Kamenshchik:2014waa,Postma:2014vaa,Weenink:2010rr,Prokopec:2012ug,Prokopec:2013zya,Domenech:2015qoa} for inflation with non-minimal coupling.}. We can perform a conformal transformation $\widetilde{g}_{\mu\nu}=\Omega^2g_{\mu\nu}$ with $\Omega^2=1-\xi\phi^2$ to go to the Einstein frame. Then the non-minimal coupling is eliminated, and we obtain a de Sitter minisuperspace model with a minimally coupled scalar field of mass $M^2=m^2+12\xi H^2$ in the small-field region of $\phi$. In this model the only constraint from the regularity requirement is that $M^2\geq 0$ \cite{Hong:2002yf}; all other constraints seem to have disappeared. The inequivalence of constraints on non-minimal coupling from Jordan and Einstein frames may seem surprising, as it contradicts the naive expectation that physics should not be changed under a conformal transformation, which is merely a change of variables. We note, however, that our minisuperspace model allows inhomogeneous fluctuations of the scalar field $\phi$, while assuming that the metric is homogeneous and isotropic. The conformal transformation $\tilde{a}^2=a^2(1-\xi\phi^2)$ would then make the new scale factor $\tilde{a}$ inhomogeneous. The new model is actually obtained by first performing the conformal transformation and then restricting to minisuperspace where the spacetime metric with the scale factor $\tilde{a}$ is homogeneous and isotropic. It is therefore not surprising that the two models are inequivalent for higher inhomogeneous modes of the scalar field. 

This argument, however, does not completely explain the inequivalence, because our constraints on $\xi$ come mostly from the homogeneous mode of the scalar field, $n=1$.  These constraints should therefore be present even in a restricted model including only the scale factor and a homogeneous scalar field.  To understand the inequivalence  in this setting, we note that the superspace boundary in the Einstein frame, ${\tilde a}=0$, corresponds to $a^2(1-\xi\phi^2) = 
a^2 - \xi\chi^2 =0$ in the Jordan frame.  For nonzero values of $\chi$ this boundary is not at $a=0$ in the original frame, so the two frames are clearly inequivalent. 
 
On the other hand, the wave function $\Psi(a,\chi_1)$ can be transformed from one frame to the other by a simple change of variables.  To see how the small-$a$ constraint on $\xi$ arises in this context, we first express the zeroth-order Einstein-frame action in terms of the original scale factor,
\begin{align}
S^\pm(\tilde{a})&=\mp\frac{1}{3H^2}\left(1-\tilde{a}^2H^2\right)^\frac32\\
&\approx \mp\frac{1}{3H^2}\left[1-\frac32a^2(1-\xi\phi^2)H^2\right]\\
&\approx S^\pm(a)\mp\frac12\xi\chi^2,
\end{align} 
where we took the limit of $a\rightarrow0$. Substituting this in the Einstein-frame wave function and assuming that $\chi$ is homogeneous, we obtain
\begin{align}
\Psi(a\approx0, \chi_1)&\approx A\exp\left[-\frac{12\pi^2}{\hbar}\left(S^\pm(a)\mp\frac12\xi\chi^2\right)-\frac{1}{2\hbar}(1\pm1)\chi_1^2\right] \\
&=A\exp\left[-\frac{12\pi^2}{\hbar}S^\pm(a)-\frac{1}{2\hbar}(1\pm1\mp6\xi)\chi_1^2\right],
\end{align}
where we have used that $\int d\Omega_3 \chi^2 = (2\pi^2)\chi^2 =\chi_1^2$. If the variable $\chi_1$ is assumed to span its original range, $-\infty<\chi_1 <\infty$, the regularity condition $1\pm1\mp6\xi\geq0$ yields the constraint $0\leq\xi\leq1/3$.

\acknowledgments
This work was supported in part by the National Science Foundation under grant PHY-1820872. MY is supported by an Allen Cormack Fellowship at Tufts University. We are grateful to the anonymous referee for very useful comments.

\appendix

\section{Mode functions $X_\lambda^n (z)$}\label{app:2ndsol}

We argued in Sec.~\ref{subsec:subdominant} that, for the tunneling wave function, the mode functions $X_\lambda^n(z)$ [Eq.~(\ref{eq:1stsol})] can be used for some low-$n$ modes if the corresponding functions ${\tilde R}_n(z)$ satisfy 
\begin{align}
\frac{\mathrm{d}}{\mathrm{d}z} {\rm Re} {\tilde R}_n(z=0) >0.
\label{d/dz2}
\end{align}
In this case the regularity condition at small $a$ cannot be consistently applied on the subdominant growing branch of the wave function and one should use only the regularity condition on the decreasing branch.  At $a\to 0$ it gives [see Eq.~(\ref{z-1})]
\begin{align}
{\tilde R}_n^+(z\to-1)=-n+(1-6\xi)\geq 0.
\label{RX}
\end{align}    
This can hold only if $\xi\leq 0$, and for $\xi>-1/6$ it can be satisfied only for the $n=1$ mode.

\begin{figure}
\includegraphics[width=0.48\textwidth]{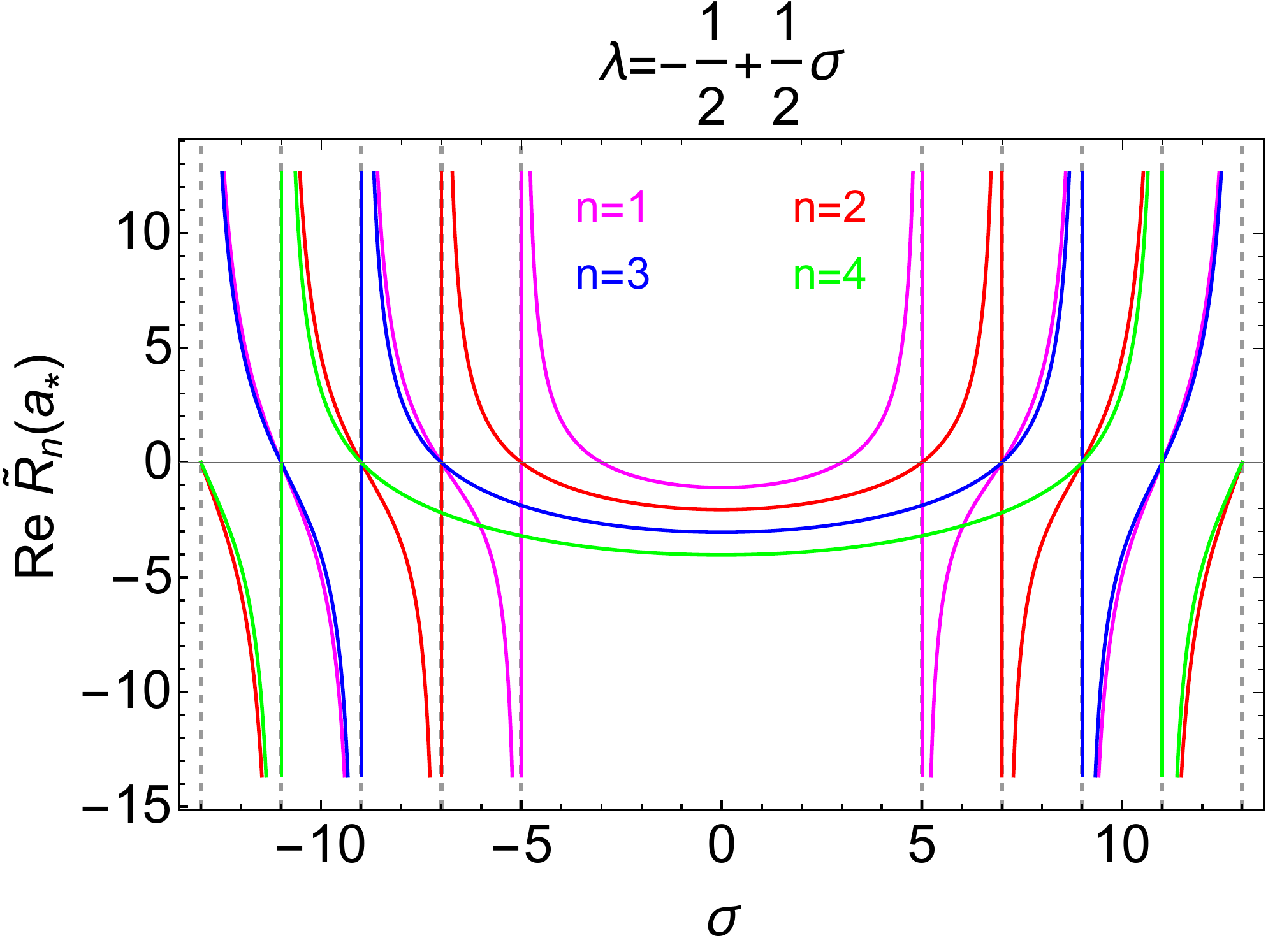}
\includegraphics[width=0.48\textwidth]{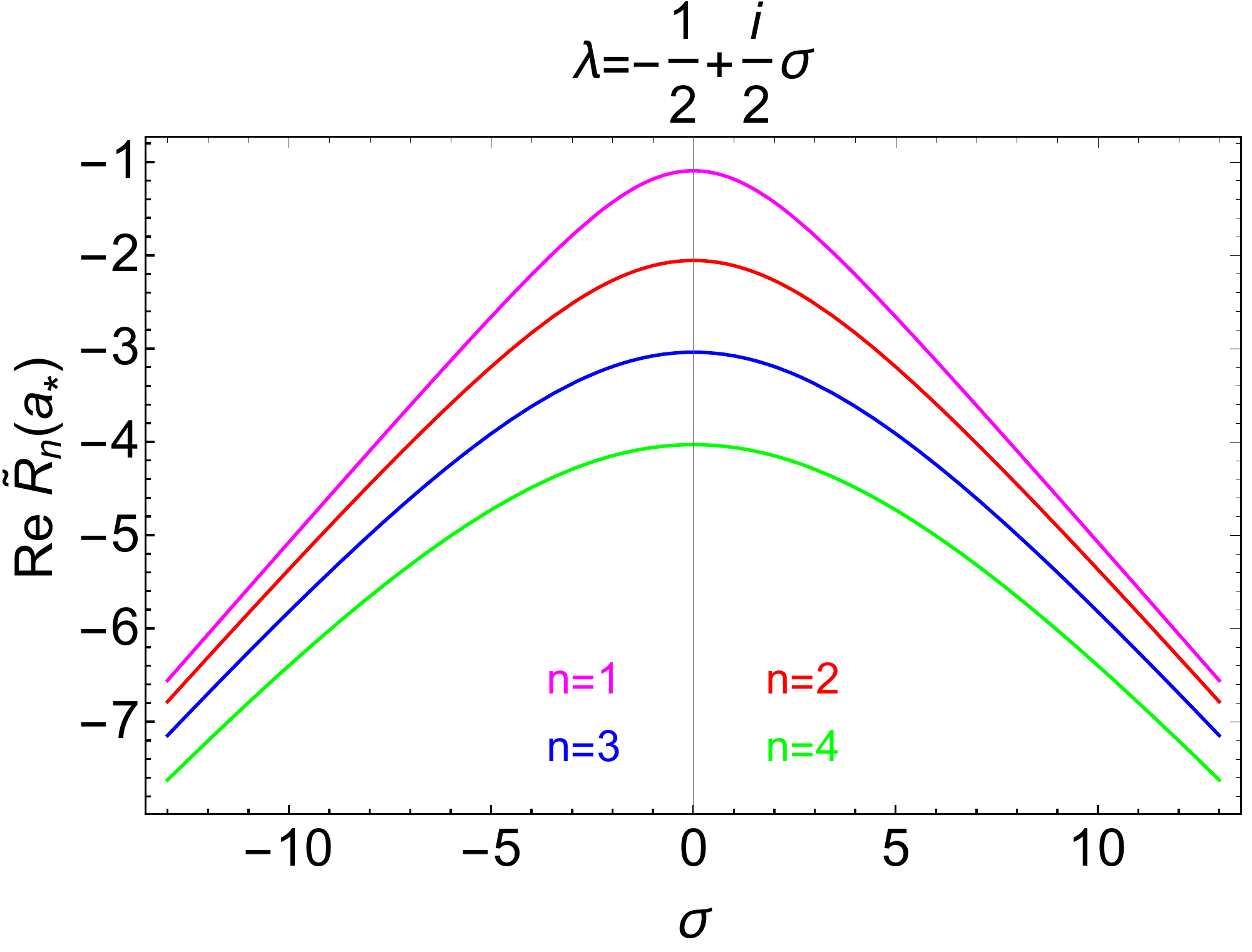}\\
\caption{The behavior of $\mathrm{Re}\tilde{R}_n(a_*)$ as functions of $\sigma$ for some low-$n$ modes.}\label{fig:ABturn}
\end{figure}

One should also check that the functions ${\tilde R}_n(z)$ satisfy the regularity condition at the turning point $a=a_*$ $(z=0)$.  In Fig.~\ref{fig:ABturn} we plot ${\rm Re} {\tilde R}_n(z=0)$ as functions of $\sigma$ for several low-$n$ modes.  The plots show that regularity at $a_*$ can be satisfied for $n=1$ only when $\sigma$ is in the range $3\leq|\sigma|\leq 5$ with $T\geq0$ (left panel). The corresponding range of the effective mass parameter $\mu^2 \equiv (m^2+\xi R)/H^2$ is $-4\leq \mu^2 \leq 0$.    
From now on we focus on the mode $n=1$.

\begin{figure}
\centering
\includegraphics[width=0.48\textwidth]{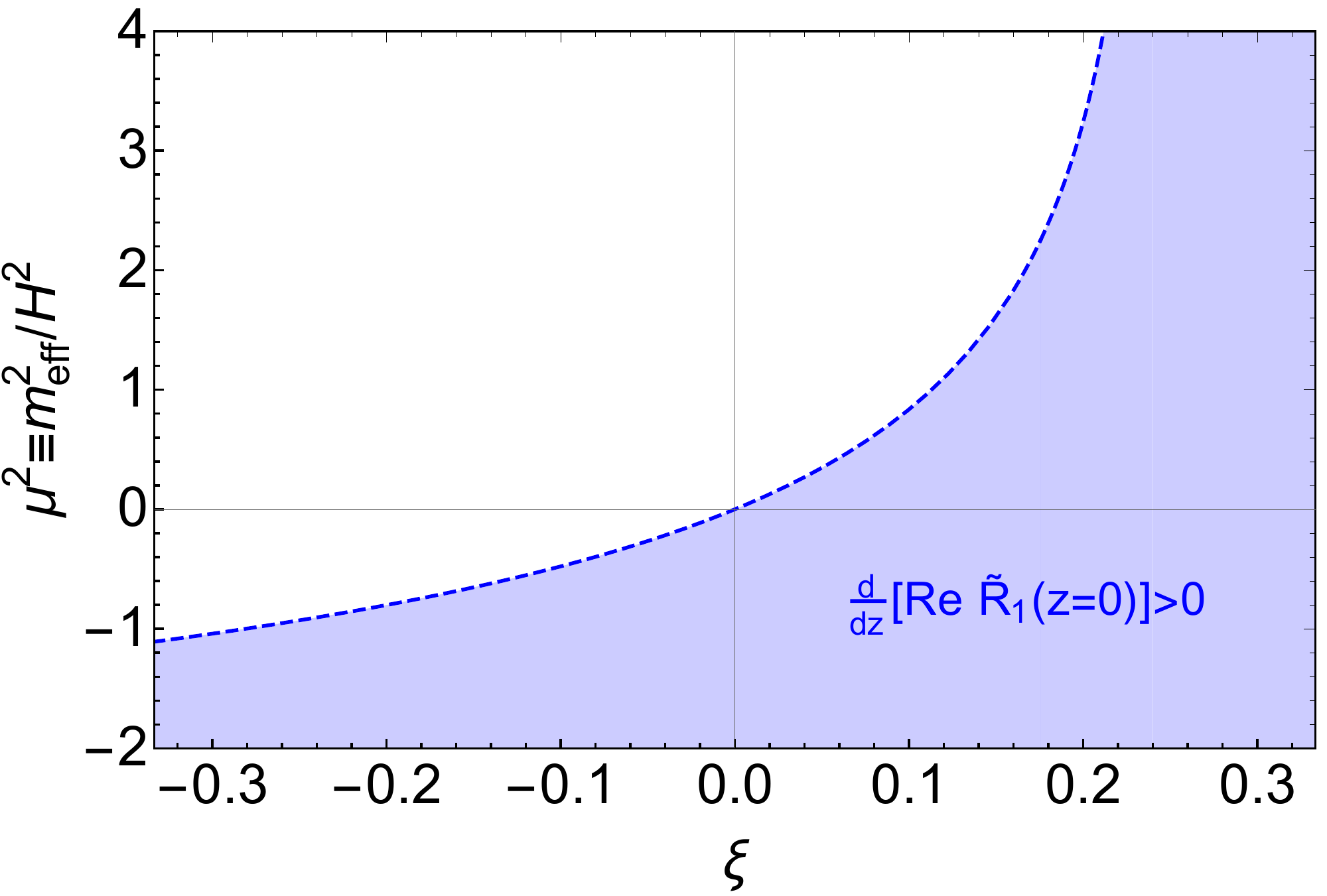}
\includegraphics[width=0.48\textwidth]{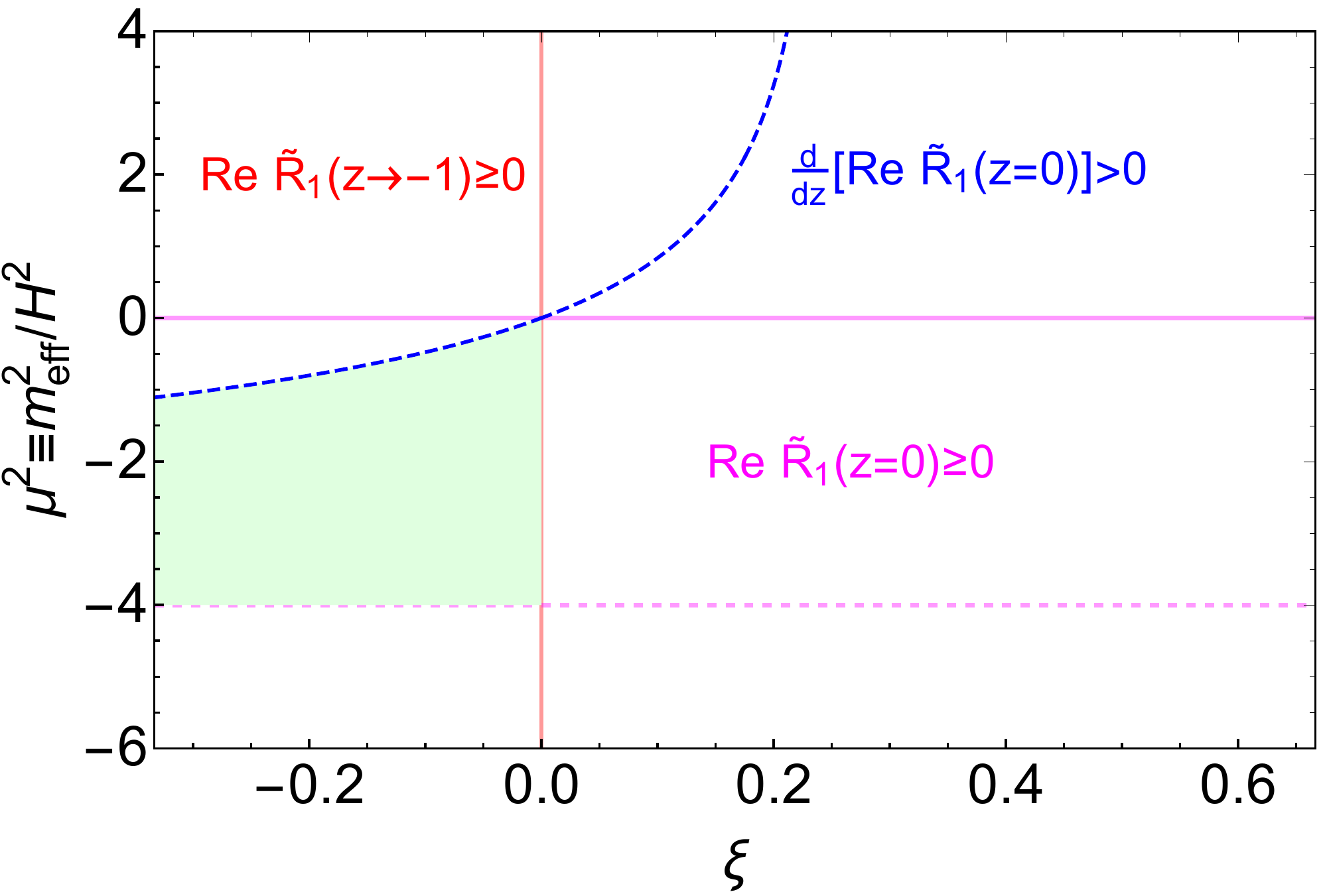}\\
\caption{The left panel shows the region (shaded in purple) where $(\mathrm{d}/\mathrm{d}z)\tilde{R}_1(z=0)>0$. In the right panel, the shaded area is an overlap of the region in left panel with the rectangular region specified by the regularity conditions at the turning point and at small $a$.}\label{fig:final}
\end{figure}

We show the region of the parameter space where the inequality in Eq.~(\ref{d/dz2}) is satisfied for $n=1$ as the shaded region in the left panel of Fig.~\ref{fig:final}.  This is the region where the mode functions $\nu_1(z)=X_\lambda^1(z)$ can potentially be used.  Furthermore, the regularity conditions for ${\tilde R}_1$ at $a\to 0$ and $a=a_*$ specify a rectangular region $\xi<0, ~ -4< \mu^2<0$.  Overlap of this latter region with the shaded region in the left panel, marked by the green shading in the right panel of Fig.~\ref{fig:final}, shows the region of the parameter space where the mode functions $X_\lambda^1(z)$ can be used for the mode $n=1$, and the constraints \eqref{eq:xibounda0} and \eqref{eq:lowerbound0} can be evaded.

\bibliographystyle{JHEP}
\bibliography{ref}

\end{document}